Unraveling electronic-structure features for metallic $Na_{0.33}CoO_2$ and charge-ordered $Na_{0.5}CoO_2$ by high-resolution electron energy-loss spectroscopy

R. J. Xiao[1], J. H. Chen[1,2], H. X. Yang[1], C. Ma[1], Q. Xu[2], H. W. Zandbergen[2], and J. Q. Li[1]*

[1] Beijing National Laboratory for Condensed Matter Physics, Institute of Physics, Chinese Academy of Sciences, Beijing, 100080, P. R. China

[2] NCHREM, Kavli Institute of Nanoscience, Delft University of Technology, Lorentzweg 1, 2628 CJ Delft, The Netherlands

Measurements by high-resolution electron energy-loss spectroscopy (HREELS) of $Na_xCoO_2$ reveal spectral features that differ remarkably between the metallic $Na_{0.33}CoO_2$ and the charge-ordered insulator $Na_{0.5}CoO_2$. Calculations by density functional theory plus Hubbard U (DFT+U) demonstrate that these differences arise essentially from the relatively greater strength of electron correlation in addition to the crystal-field effect in $Na_{0.5}CoO_2$. The effective U values are estimated to be ~3.0eV for x=0.5 and ~1.0eV for x=0.33, respectively. The electronic structures for these correlation strengths give good interpretations for the physical properties observed in the materials.

Recently, the $Na_xCoO_2$ system has been extensively studied following the discovery of superconductivity in $Na_{0.33}CoO_2 \cdot 1.3H_2O$ [1-4], and a rich variety of notable physical properties observed in this system are now believed to be connected with the presence of electron correlation [5-7], such as the charge ordering phenomenon observed in $Na_{0.5}CoO_2$ [8]. Actually, strongly correlated materials constitute a wide class of materials in which the electrons are strongly correlated so that the behavior of electrons cannot be well described by simple one-electron theories [9,10]. Many transition metal oxides belong to this class, including high-temperature superconductors [11,12] and colossal magnetoresistant manganites [13,14]. The strength of the on-site Coulomb interaction among the Co $d$ electrons in $Na_xCoO_2$, as well as the electronic structure, could vary with the doping level due to the change in electron screening [15,16]. Metallic $Na_{0.33}CoO_2$, becomes a superconductor with $H_2O$ intercalation [1], while charge ordered $Na_{0.5}CoO_2$, on the other hand, becomes an insulator below 50K [8]. Such observations of extremely different properties of materials that are made up of rather similar units suggest notable alterations in their electronic structures from x=0.3 to 0.5. Though certain previous theoretical studies based on first principle calculations examined the electronic structures with correlation effects in this layered system [15,16], experimental measurements directly revealing the fine features of electronic structures are very much needed for a fundamental understanding of the specific properties of these materials. Recently developed TEM instruments implemented with a monochromator can record the HREELS data with an energy resolution of ~0.1eV, which allows us to explore the fine electronic structures of a variety of systems. In this paper, the experimental HREELS data from the metallic $Na_{0.33}CoO_2$ and the charge-ordered $Na_{0.5}CoO_2$ are extensively analyzed by the DFT+U method, and the electronic structure, electron correlation strength and notable physical properties are discussed in comparison with previously reported results. It is also pointed out that determination of fine electronic structure based on HREELS performed in a modern TEM in combination with DFT+U calculations could be a unique method that will allow us to better understand the electronic properties of nanostructured materials and devices of strongly correlated systems.

The single-crystalline $Na_{0.5}CoO_2$ samples used in the present study were prepared by sodium deintercalation from $Na_{0.85}CoO_2$ grown by the traveling-solvent floating zone method. The sodium content in our samples was determined by the inductively-coupled plasma atomic-emission spectroscopy (ICP) as reported in our previous papers [17]. The specimens for transmission electron microscopy (TEM) observations were prepared simply by crushing the bulk material into fine fragments, which were then supported by a copper grid coated with thin carbon film. TEM observations and HREELS analysis were carried out in a Tecnai F20 TEM instrument implemented with a monochromator and equipped with a post-column Gatan imaging filter, operating at 200 KV. The achievable energy resolution for HREELS is ~0.1eV.

In our theoretical calculations, we used the full potential linearized augmented-plane-wave+local orbital (APW+LO) model as implemented in WIEN2k code [18], in the generalized gradient approximation (GGA). For $Na_{0.5}CoO_2$, the orthorhombic cell with space group *Pmmn* was used as determined by powder neutron diffraction, in which the ordered Na ions form one-dimensional zigzag chains, resulting in two types of Co sites with different bond lengths to oxygen atoms [8]. In our analysis, $R_{mt}K_{max}$ was set to 7.0 to determine the basis size. The Hubbard parameter U was considered from 0 to 5.0eV. The EELS spectral simulations were performed using the newly updated TELNES.2 package [19] of the WIEN2k distribution.

Figure 1(a) shows two instances of experimental EELS data of the oxygen $K$ edge of $Na_{0.5}CoO_2$ obtained from the same sample, the red curve recorded with the monochromator has a high resolution of ~0.2eV, and the blue one has a conventional resolution of ~0.9eV recorded without the monochromator. The most striking feature revealed by HREELS is the presence of fine spectral features in detail, especially the splitting of peak *a* yielding two clear sub-peaks, 1eV apart. These peaks are essentially connected with the electron transitions from the oxygen 1$s$-state to the vacant states above the Fermi level ($E_F$). In order to understand the electron excitations responsible for the fine spectral features, we performed the DFT+U calculations with U values ranging from 0 to 5.0eV. Figure 1(b) shows the theoretical high-resolution spectra of peak *a* in the oxygen $K$ edge of $Na_{0.5}CoO_2$ with a resolution of 0.2eV. It is notable that the spectrum for U=0 to 2.5eV has a single peak without visible splitting, and this peak shifts toward higher energy with an increase of U. As U becomes larger than 2.5eV, the splitting of peak *a* starts. We found that the theoretical spectrum for U=3.0eV accurately matches with the fine details of the experimental HREELS data (Fig. 1(a)); i.e., the 1eV splitting of peak *a* is well reproduced.

Figure 2(a) shows an experimental HREELS of $Na_{0.33}CoO_2$ with the energy resolution of 0.2eV. It is generally agreed that the Coulomb correlation should be considered for better understanding of electronic structure and physical properties of $Na_{0.33}CoO_2$ and its superconducting hydration $Na_{0.33}CoO_2 \cdot 1.3H_2O$ (Tc~5K) [15,16,20], so we carried out a calculation using the DFT+U method to simulate the HREELS data with the correlation strength ranging from 0 to 5.0eV. Figure 2(b) shows a series of the simulated O-$K$ edge for $Na_{0.33}CoO_2$ with 0≤U≤5.0eV. The DFT+U calculations for $Na_{0.33}CoO_2$ were performed with the space group P6$_3$/mmc using the virtual crystal approximation [21]. It is demonstrated that an increase of U pushes the peak toward higher energy. Our further analysis suggests that the experimental O *K* edge has a symmetrical σ-functional feature, therefore the peak *a* is likely to have a completely symmetrical profile. The theoretical spectrum of U=1.0eV shows structure very similar to the experimental data, which suggests that the effective Coulomb value U in Na$_{0.33}$CoO$_2$ is smaller than that in Na$_{0.5}$CoO$_2$ as discussed above.

According to the determined U values, we performed an analysis of the electronic structures for Na$_{0.5}$CoO$_2$ with U=3.0eV and Na$_{0.33}$CoO$_2$ with U=1.0eV. The results for Na$_{0.33}$CoO$_2$ are in good agreement with the data reported in previous literature in which the metallic conductivity and other related physical properties were systemically discussed [15]. Figure 3 shows the density of states (DOS) of Na$_{0.5}$CoO$_2$ calculated for U=3.0eV, illustrating the notable effect of electron correlation on the band structure. It is clearly visible in the inset figure that a narrow energy gap of ~0.056eV appears at the Fermi level, fundamentally in agreement with the photoemission measurements [22]. The presence of this energy gap directly demonstrates an insulating ground state for Na$_{0.5}$CoO$_2$ as was commonly discussed in previous publications [8,16]. Moreover, the theoretical results for U=3.0eV also show that the Co e$_g$ band at about 1.5eV above E$_F$ is divided into two main manifolds separating from each other by ~1.0eV – the first one is from 1.5 to 2.5eV and the other one is from 2.5 to 3.5eV. These two manifolds correspond to the hybridization states O 2*p* with Co2 3*d* and Co1 3*d* orbitals, respectively. Actually, the double sub-peaks of peak *a* in the HREELS (Fig. 1(a)) are directly connected with the splitting of electronic states between Co1 and Co2. The first subpeak of peak *a* actually originates from the excitation from O 1*s* to O 2*p* orbital hybridized with Co2 e$_g$;

the second subpeak about 1eV higher is from the transition from O 1$s$ to the hybridized states of O 2$p$ and Co1 $e_g$.

The existence of electron correlation of U=3.0eV could also explain the HREELS data for the cobalt $L_{2,3}$ edge in Na$_{0.5}$CoO$_2$. For cobalt, a transition metal, the $L_{2,3}$ edge in EELS corresponds with the electron excitations from Co 2$p^{3/2}$ and 2$p^{1/2}$ states to the empty states above $E_F$. Figure 4(a) shows the experimental and theoretical HREELS of the Co-$L_{2,3}$ edge for Na$_{0.5}$CoO$_2$. It is noted that our experimental spectrum shows very similar features with the X-ray absorption spectra (XAS) as reported in ref.[23] – peak *a* (~781eV) and peak *b* (~794eV) are respectively assigned to Co 2$p^{3/2}$ and 2$p^{1/2}$. Our theoretical simulation demonstrates that the excitations of Co2 and Co1 sites could yield slightly different features in the EELS as illustrated in the bottom curve – the peak *b* therefore consists actually of two sub-peaks corresponding respectively to the excitations of Co2 and Co1 sites. It is also noted that the calculated data for peak *a* certainly has structures differing from the experimental results. Similar structural contrast was observed in the interpretation of the XAS experimental data based on the theoretical analysis of electronic states of the Co$^{4+}$O$_6$ and Co$^{3+}$O$_6$ clusters [24]. Actually, this kind of discrepancy originates chiefly from the redistribution of spectral weight caused by the core-hole effect and/or spin-orbit coupling [25]. These factors in the Na$_x$CoO$_2$ system could reduce the spectral intensity from Co$^{3+}$O$_6$ cluster [26]. Our analysis suggests that the theoretical contribution of Co2 is more than two times as large as is suggested by experimental data. Thus we show a corrected spectrum of Na$_{0.5}$CoO$_2$ by reducing the intensity ratio between Co1 and Co2 which gives rise to a corrected curve in agreement with the experimental results. M.X. Labute *et al.* [26] pointed out that a spectral weight shift often appears in systems where the mixed valence state is associated with Co ions. The spectral weight shift toward the high spin $Co^{4+}$ configuration can be explained based on the core-hole interaction of the 3$d$-electrons. In order to simulate the core-hole effect in $Na_{0.5}CoO_2$ by WIEN2k code, we must construct a large supercell to overcome the effects of interaction among core-holes in neighboring cells. Our further study considering the electronic structure of the excited states is still underway.

Electron correlation strength of U=3.0eV is sufficiently strong to cause charge disproportionation and spin ordering [16]. Our further calculations suggest that the appearance of the band gap in $Na_{0.5}CoO_2$ is accompanied by the spin ordering of the Co1 and Co2 ions. Co1 and Co2 are two distinct sites in the orthorhombic $Na_{0.5}CoO_2$ structure which form parallel zigzag atomic chains along the <110> direction. The magnetic moments for Co1 and Co2 with U=3.0eV are 0.08$\mu_B$ and 0.72$\mu_B$ respectively. These results are essentially consistent with the data of K.-W. Lee *et al.* [16]. They performed the calculations based on a proposed supercell with a space group *P2/m* and suggested the possible occurrence of charge disproportionation in connection with the moderate correlation effect in this material. Figure 4(b) presents the calculated electron spin density in $Na_{0.5}CoO_2$ with U=3.0eV, demonstrating that the electron localization occurs mainly on the $dz^2$ orbitals of Co2 ions.

In summary, the HREELS measurements on spectral features of the $Na_xCoO_2$ materials demonstrate remarkably different structures between the metallic $Na_{0.33}CoO_2$ and the charge-ordered insulator $Na_{0.5}CoO_2$. This difference can be well understood by the existence of notable electron correlation (~3.0eV) in $Na_{0.5}CoO_2$ as illustrated by DFT+U calculations. The electronic structure for U=3.0eV gives good interpretations for the observed physical properties in the x=0.5 material, such as charge disproportionation and spin ordering. Results for $Na_{0.33}CoO_2$ with a relative weak electron correlation reveal clear metallic properties in good agreement with the experimental results. It is worth pointing out that HREELS performed in a modern TEM implemented with a monochromator is capable of probing electronic structures with an energy resolution of ~0.1eV from an area as small as a few angstroms in size. Therefore, HREELS combined with the DFT+U calculation could be a uniquely effective method that will allow us to better understand the fine electronic structures of nanostructured materials and devices of strongly correlated systems.

**Acknowledgments**

The work reported here is supported by the National Science Fund of China (grant No. 10474102) and by the Ministry of Science and Technology of China (973 project No: 2006CB601001).

Figure captions

Fig. 1 (a) Experimental HREELS for O $K$ edge in $Na_{0.5}CoO_2$. The spectrum obtained without monochromator is shown for comparison. (b) Simulated spectra for O $K$ edge with $0 \leq U \leq 5.0 eV$.

Fig. 2 (a) Experimental HREELS for O $K$ edge in $Na_{0.33}CoO_2$. (b) Simulated spectra for O $K$ edge with $0 \leq U \leq 5.0 eV$.

Fig. 3 Density of states for $Na_{0.5}CoO_2$ with $U=3.0eV$. An energy gap is clearly indicated in the inset.

Fig. 4 (a) Experimental and calculated HREELS of Co-$L_{2,3}$ edge in $Na_{0.5}CoO_2$. The corrected spectrum was obtained by adjusting the intensity ratio between the transitions from Co1 and Co2 ions to fit the experimental spectrum. (b) Upper image: schematic model for $CoO_2$ layer; Lower image: isosurface of the electron-spin-density obtained by subtracting the spin-down density from the spin-up density with $U=3.0eV$. The $dz^2$-orbital order is shown.

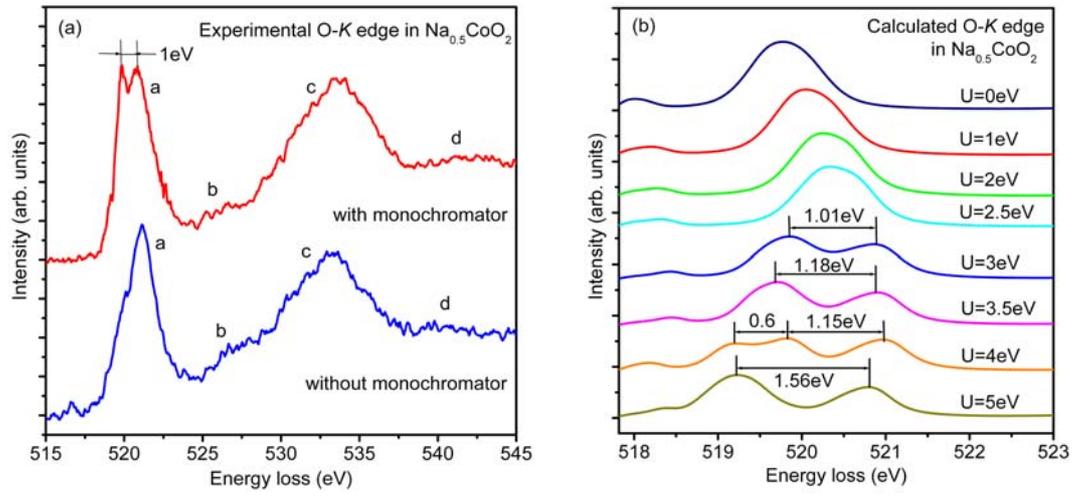

Fig. 1

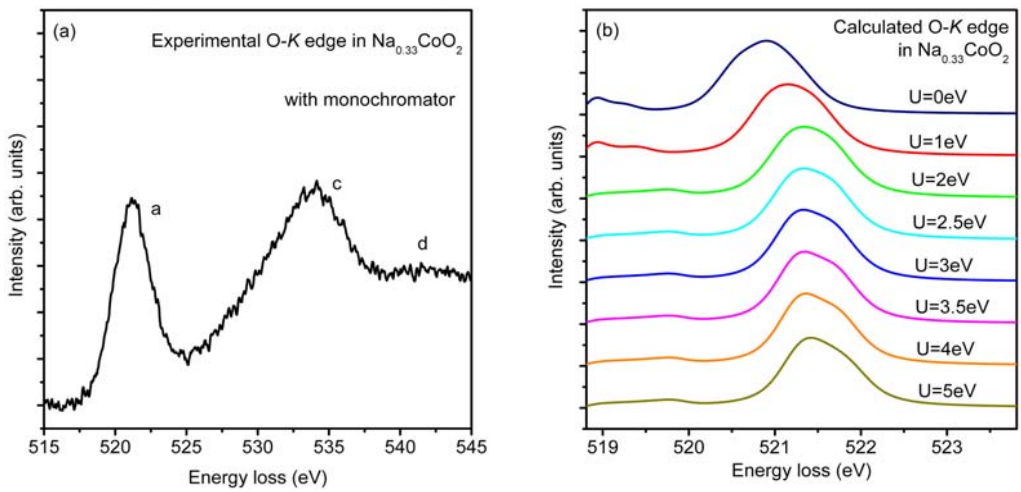

Fig. 2

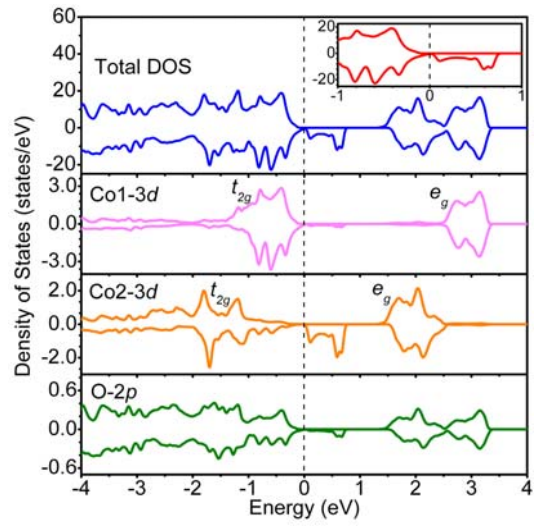

Fig. 3

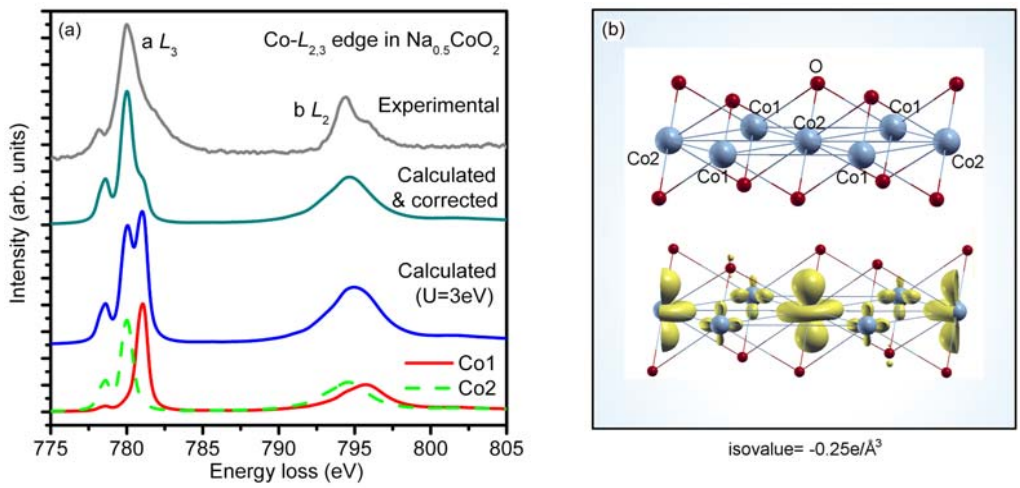

Fig. 4